\documentclass[letterpaper,twocolumn,10pt]{article}
\usepackage{usenix,epsfig,endnotes}

\usepackage{graphicx}
\usepackage{subcaption}
\usepackage{xspace}
\usepackage{amsmath}
\usepackage{url}

\usepackage{hyperref}

\newcommand{\smart}[1][]{\textsc{BFT-SMaRt}\xspace}
\newcommand{\wheat}[1][]{\textsc{WHEAT}\xspace}

\begin{document}

\title{A Byzantine Fault-Tolerant Ordering Service for the\\ Hyperledger Fabric Blockchain Platform}
\author{\ \ \ \ \ \ Jo\~{a}o Sousa \ \ \ \ \ \ \ \ \ Alysson Bessani \ \ \ \ \ \ \ \ \ \ \ \ \ \ \ \ Marko Vukoli\'c\\
{\small LaSIGE, Faculdade de Ciências, Universidade de Lisboa} \ \ \ \ \ \ \ \ 
{\small IBM Research Zurich}
}

\date{}

\maketitle

\begin{abstract}
Hyperledger Fabric (HLF) is a flexible permissioned blockchain platform designed for business applications beyond the basic digital coin addressed by Bitcoin and other existing networks.
A key property of HLF is its extensibility, and in particular the support for multiple ordering services for building the blockchain.
Nonetheless, the version 1.0 was launched in early 2017 without an implementation of a Byzantine fault-tolerant (BFT) ordering service.
To overcome this limitation, we designed, implemented, and evaluated a BFT ordering service for HLF on top of the \smart state machine replication/consensus library, implementing also optimizations for wide-area deployment.
Our results show that HLF with our ordering service can achieve up to ten thousand transactions per second and write a transaction irrevocably in the blockchain in half a second, even with peers spread in different continents.
\end{abstract}

\section{Introduction}

The impressive growth of Bitcoin and other blockchain platforms based on the Proof-of-Work (PoW) technique, made evident the performance limitations of this approach.
These limitations are mostly related with perfomance: existing systems are capable of processing from 7 (Bitcoin) to 10s-100s transactions per second and present transaction confirmation latencies of up to one hour~\cite{Vukolic2016}.
Several alternative blockchain platforms proposed in the last years try to avoid these limitations by employing more traditional Byzantine Fault-Tolerant (BFT) consensus protocols (e.g.,~\cite{Cas02}) for establishing consensus on the blocks in a blockchain~\cite{cachin2017}.

Hyperledger Fabric\footnote{\url{https://www.hyperledger.org/projects/fabric}.} (HLF) is a platform that target business applications.
It is built with flexibility and generality as key design concerns, supporting thus a wide variety of non-deterministic smart contracts (here called chaincode) and pluggable services~\cite{Vukolic2017}.
The support for pluggable components, gives the HLF an unprecedented level of extensibility, and in particular the support for multiple ordering services for writing transactions on the blockchain.
Despite of that, the version 1.0 (launched in early 2017) comes without any Byzantine fault-tolerant (BFT) ordering service, supporting only crash tolerance through an ordering service based on Apache Kafka.\footnote{\url{https://kafka.apache.org/}.}

In this paper, we describe our efforts in overcoming this limitation, by presenting the design, implementation, and evaluation of a BFT ordering service for HLF 1.0\footnote{PBFT implementation present in HLF v0.6 was deprecated with transition to the new v1 architecture~\cite{Androulaki2016}.} based on the \smart state machine replication/consensus library~\cite{Bessani2014}, and its extensions to support low-latency consensus on the internet~\cite{Sousa2015}.
Our preliminary evaluation, both on a local cluster and in a geo-distributed setting, show that HLF with \smart ordering service can achieve up to 10k representative transactions per second and write a transaction irrevocably in the blockchain in half a second, even with consensus nodes spread through different continents.
The source code of the our service is freely available on the internet for the HLF community.\footnote{\url{https://github.com/jcs47/hyperledger-bftsmart}.}

As an additional contribution, the paper also discuss the key concerns that need to be addressed to apply existing (BFT or not) state machine replication protocols to blockchain platforms like HLF, and the service model and workload of interest in this kind of systems, which are substantially different from the microbenchmarks~\cite{Cas02} and the Zookeeper-like client-server model~\cite{Hun10} still used to evaluate BFT protocols.

The rest of this paper is organized as follows.
We start by presenting the fundamentals of blockchain technology (Section \ref{sec:blockchain-tech}) and Hyperdeger Fabric (Section \ref{sec:hlf}). After that, the \smart and \wheat protocols (Section \ref{sec:bftsmart}) are briefly described, and we proceed to present the \smart ordering service (Section \ref{sec:smart-os}) and its experimental evaluation (Section \ref{sec:smart-os-eval}).
We discuss some related work in Section \ref{sec:rel-work} and conclude this paper in Section \ref{sec:conclusions}.

\section{Blockchain Technology}
\label{sec:blockchain-tech}

A blockchain is an open database that maintains a distributed ledger typically deployed within a peer-to-peer network.
It is comprised by a continuously growing list of records called \emph{blocks} that contain transactions \cite{Nakamoto_bitcoin:a}.
Blocks are protected from tampering by cryptographic hashes and a consensus mechanism.
%The hashes prevent the whole chain from being tampered with, while the consensus mechanism controls which blocks get to be appended to the ledger next.

\begin{figure}[!t]
\centering
\includegraphics[scale=0.4]{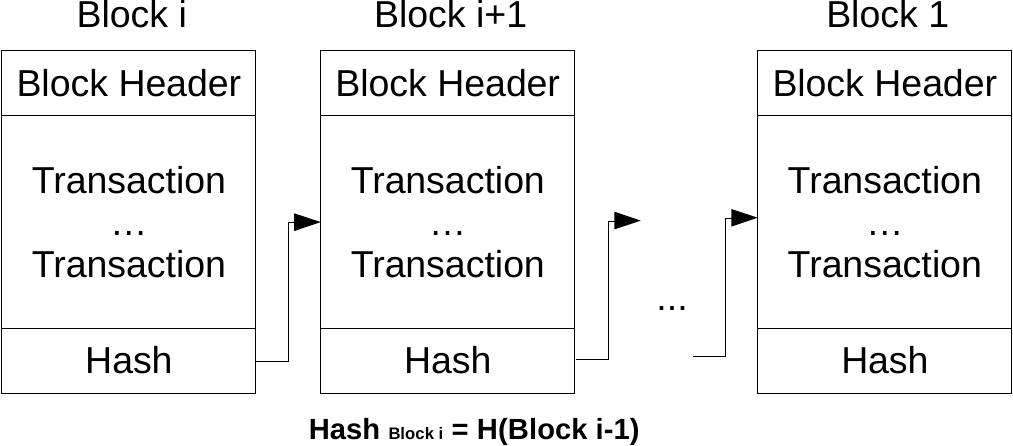}
\caption{Blockchain structure.}
\label{fig:hashchain}
\end{figure}

The structure of a blockchain -- illustrated in Figure \ref{fig:hashchain} -- consists of a sequence of blocks in which each one contains the cryptographic hash of the previous block in the chain.
This introduces the property that block $j$ cannot be forged without also forging all subsequent blocks $j+1...i$.
%Hence, all transactions up to block $N$ become immutable.
Furthermore, the consensus mechanism is used to (1) prevent the whole chain from being modified; and (2) decide which block to be appended to the ledger.

The blockchain may abide to either the \emph{permissionless} or \emph{permissioned} models \cite{Vukolic2016}.
Permissionless ledgers are maintained across peer-to-peer networks in a totally decentralized and anonymous manner\cite{Nakamoto_bitcoin:a, Buterin2015}.
In order to determine which block to append to the ledger next, peers need to execute Proof-of-Work (PoW) consensus \cite{Garay2015}.
The key idea behind PoW consensus is to limit the rate of new blocks by solving a cryptographic puzzle, i.e., execute a CPU intensive computation that takes time to solve, but can be verified quickly.
This is achieve by forcing peers to find a nonce $\mathit{N}$ such that given their block $\mathit{B}$ and a limit $\mathit{L}$, the cryptographic hash of $\mathit{B||N}$ is lower than $\mathit{L}$ \cite{Back02hashcash, Dwork:1992:PVP:646757.705669}.
The first peer that presents such solution gets its block appended to the ledger.
Roughly speaking, as long as the adversary controls less than half of the total computing power present in the network, PoW consensus prevents the adversary from creating new blocks faster than honest participants.
%Hence, it would be unfeasible to catch-up without obtaining the majority of the computing power present in the network.

Permissionless blockchains have the benefit of enabling the ledger to be curated completely anonymously; any peer willing to hold a copy of the ledger and create new blocks to it is able to do so.
On the other hand, the computational effort associated to PoW consensus is both energy- and time-consuming; even if specialized hardware is used to find a Proof-of-Work, this process still exerts a limit on transaction latency.

By contrast, permissioned blockchains require a set of trusted nodes tasked with creating new blocks and executing a traditional Byzantine consensus protocol to decide the order by which the blocks are inserted to the ledger \cite{Kwon2016, Martino2016, Vukolic2017, cachin2017}.
%Instead of finding and including a nonce in each block, the nodes include a digital signature instead.
Hence, permissioned blockchains do not expend the amount of resources that open blockchains do and are able to reach better transaction latency and throughput.
%This is the blockchain model in which large portions of the industry are currently more interested in: permissioned blockchains not only exhibit better performance, but also make it possible to control the set of participants tasked with maintaining the ledger -- thus separating it from the dark web or illegal activities.
In addition, it makes possible to control the set of participants tasked with maintaining the ledger -- rendering this type of blockchain a more attractive solution for larger corporations, since it can be separated from the dark web or illegal activities.

\section{Hyperledger Fabric}
\label{sec:hlf}

Hyperledger Fabric (HLF) \cite{Vukolic2017, Cachin2016} is an open-source project within the Hyperledger umbrella project.\footnote{\url{https://www.hyperledger.org/}}
It is a modular permissioned blockchain platform designed to support pluggable implementations of different components, such as the ordering and membership services.
HLF enables clients to manage transactions by using chaincodes, endorsing peers and an ordering service.

Chaincode is HLF's counterpart for smart contracts \cite{Szabo1996}.
It consists of code deployed on the HLF's network, where it is executed and validated by the endorsing peers, who maintain the ledger, the state of a database (modeled as a versioned key/value store), and abide by endorsement policies.
The ordering service is responsible for creating blocks for the distributed ledger, as well as the order by which each blocks is appended to the ledger.

\paragraph{HLF protocol.}

The HLF general transaction processing protocol~\cite{Androulaki2016} -- depicted in Figure \ref{fig:fabric-protocol} --  works as follows:

\begin{figure*}[!t]
\centering
\includegraphics[page=2,scale=0.6]{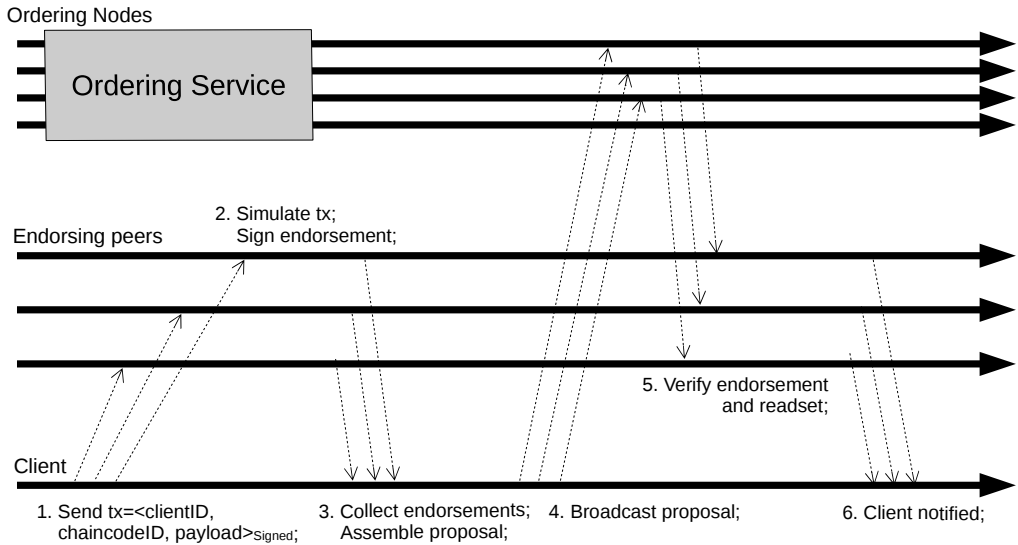}
\caption{HLF protocol.}
\label{fig:fabric-protocol}
\end{figure*}
\begin{enumerate}

\item \emph{Clients create a transaction and send it to endorsing peers}. This message is a signed request to invoke a chaincode function. It must include the chaincode ID, timestamp and the transaction's payload.

\item \emph{Endorsing peers simulate transactions and produce an endorsement signature}. They must verify if the client is properly authorized to perform the transaction by evaluating access control policies of a chaincode. Transactions are then executed against the current state. Peers transmit to the client the result of this execution (read and write sets associated to their current state) alongside the endorsing peer's signature. No updates are made to the ledger at this point.

\item \emph{Clients collect and assemble endorsements into a transaction.} The client verifies the endorsing peers signatures, determine if the responses have the matching read/write set and checks if the endorsement policies has been fulfilled. If these conditions are met, the client creates a signed envelope with the peers' read and write sets, signatures and the Channel ID.\footnote{A channel is a private blockchain on a HLF network, providing data partition. Each peers of the channel share a channel-specific ledger.} The aforementioned envelope represents a \emph{transaction proposal}.

\item \emph{Clients broadcast the transaction proposal to the ordering service.} The ordering service does not read the contents of the envelope; it only gathers envelopes from all channels in the network, orders them using atomic broadcast, and creates signed chain blocks containing these envelopes.

\item \emph{The blocks of envelopes are delivered to the peers on the channel.} The envelopes within the block are again validated to (1) ensure the endorsement policies were fulfilled, and (2) to check if there were changes to the peers' state for read set variables (since the read set was generated by the transaction execution). To this end, the read set contains a set of versioned keys that endorsing peers read at the time of simulating a transaction (step 2). Depending on the success of these validations, the transaction proposal contained in envelopes are marked as either being valid or invalid.

\item \emph{Peers append the received block to the channel's blockchain.} For each valid transaction, the write sets are committed to the peers' current state. An event is triggered to notify the client that the transaction has been immutably appended to the channel's blockchain, as well as notification of whether the transaction were deemed valid or invalid. Notice that invalid transactions are also added to the ledger, but they are not executed at the peers. This also has the added benefit of making it possible to identify malicious clients, since their actions are also recorded on the ledger.
\end{enumerate}

An important aspect of the HLF protocol is that endorsement (step 2) and validation (step 5) can be done at different peers.
Furthermore, contrary to the chaincode execution during endorsement, the validation code needs to be deterministic, i.e., the same transaction validated by different peers in the same state produces the same output \cite{Vukolic2017}.

\paragraph{Pluggable consensus.}

As mentioned before, HLF is a modular blockchain platform.
In particular, one of the components that support plug-and-play capability is the ordering service.
Currently, HLF's codebase includes the following ordering service modules: (1) a centralized, non-replicated ordering service that does not execute any distributed protocol that is used mostly for testing the platform; and (2) a replicated ordering service capable of withstanding crash faults, consisting of an Apache Kafka cluster\footnote{\url{https://kafka.apache.org/}} and its respective ZooKeeper ensemble \cite{Hun10}.
At the time of this writing, both modules have limitations.
The non-replicated module requires very few hardware resources, but it is also a single point of failure.
The Kafka-based module is both decentralized and robust, but can only withstand crash faults.

%SimpleBFT: Replicated ordering service capable of withstanding Byzantine faults, consisting of a modified implementation of the PBFT protocol \cite{Cas02}.
%SimpleBFT is still in development and is not supported in the current stable version of HLF (1.0).

\section{\smart \& \wheat}
\label{sec:bftsmart}

The ordering service presented in this paper was designed on top of existing BFT systems, namely \smart~\cite{Bessani2014} and \wheat~\cite{Sousa2015}. 
In this section we present a brief description of these works.

\smart implements a modular SMR protocol on top of a Byzantine consensus algorithm~\cite{Sousa12:::}.
Under favourable network conditions and the absence of faulty replicas \smart executes the message pattern depicted in Figure \ref{fig:bftsmart-pattern}, which is similar to the PBFT protocol~\cite{Cas02}.

Clients send their requests to all replicas, triggering the execution of the consensus protocol.
%Depending on \smart's configuration, the requests can either contain a MAC vector (to withstand non-malicious faults) or a digital signature (to tolerate malicious clients).
Each consensus instance $i$ begins with one replica -- the \emph{leader} -- proposing a batch of requests to be decided within that consensus.
This is done by sending a PROPOSE message containing the aforementioned batch to the other replicas.
All replicas that receive the PROPOSE message verify if its sender is the leader and if the batch proposed is valid.
If this is the case, they register the batch being proposed and send a WRITE message to all other replicas containing a cryptographic hash of the proposed batch.
If a replica receives $\lceil\frac{n+f+1}{2}\rceil$ WRITE messages with the same hash, it sends an ACCEPT message to all other replicas containing this hash.
If some replica receives $\lceil\frac{n+f+1}{2}\rceil$ ACCEPT messages for the same hash, it deliver its correspondent batch as the decision for its respective consensus instance.

This is the message pattern that is executed if the leader is correct and the system is synchronous.
If these conditions do not hold, the protocol needs to elect a new leader and force all replicas to converge to the same consensus execution.
This mechanism is dubbed \emph{synchronization phase} and is described in detail in \cite{Sousa12:::}.
We do not describe it in this work because our experiments do not evaluate this part of the protocol.

Our ordering service also employs \wheat, a variant of \smart optimized for geo-replicated environments.
It differs from the aforementioned protocol in the following way: it employs the tentative executions proposed in~\cite{Cas02} and uses a vote assignment scheme for efficient quorum usage introduced in \cite{Sousa2015}.
Tentative execution consists of delivering client requests right after finishing the WRITE phase, thus executing the ACCEPT phase asynchronously. %\footnote{In PBFT these communication steps are named PREPARE and COMMIT.}
This optimization comes at the cost of (1) potentially needing to perform a rollback on the application state if there is a leader change, and (2) forcing clients to wait for $\lceil\frac{n+f+1}{2}\rceil$ messages from replicas (instead of $f+1$)~\cite{Cas02}.
Moreover, the vote assignment schemes integrate the classical ideas of weighted replication~\cite{Garcia-Molina85,Gif79,Paris86} to state machine replication protocols by relying primarily on the fastest replicas present in the system while still preserving its original safety and liveness properties of the protocol.
This mechanism improves latency by allowing more choice: if there is a spare replica in the system that is faster than the rest, the optimal quorum will contain this replica.

\begin{figure}[t]

%		\centering
						
%    \begin{subfigure}{.49\textwidth}
        \includegraphics[page=3,scale=0.85]{figs/smart-msgpattern.pdf}
				\caption{\smart message pattern.}
        \label{fig:bftsmart-pattern}
%    \end{subfigure}
%    
%    \begin{subfigure}{.4\textwidth}
%        \includegraphics[page=7,scale=0.63]{figs/smart-msgpattern.pdf}
%				\caption{\wheat message pattern.}
%        \label{fig:wheat-pattern}    
%    \end{subfigure} 
%		
%		
%    \caption{\smart message pattern during fault-free executions.}
%		\label{fig:message-pattern}
%
\end{figure}

\section{BFT-SMaRt Ordering Service}
\label{sec:smart-os}

\begin{figure}[t]
\centering
\includegraphics[width=\columnwidth]{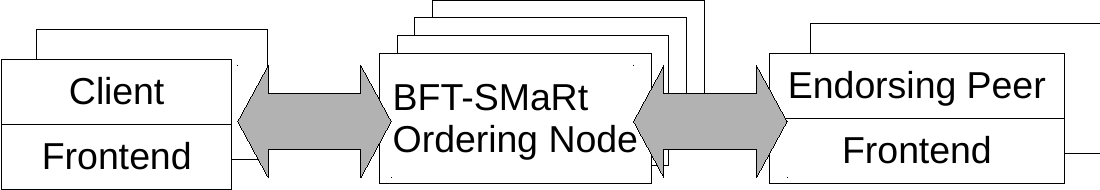}
\caption{BFT-SMaRt ordering service.}
\label{fig:bftsmart-orderer}
\end{figure}

The BFT-SMaRt module for HLF's ordering service consists of an ordering cluster and a set of frontends (Figure  \ref{fig:bftsmart-orderer}).
The ordering cluster is composed by a set of $3f + 1$ nodes that collect envelopes from the frontends and execute the \smart's replication protocol with the purpose of totally ordering these envelopes among them.
Once a node gathers a predetermined number of envelopes, it creates a new block containing these envelopes and a hash of the previously created block, generates a digital signature for the block, and disseminates it to all known frontends, which collect $2f+1$ matching blocks from ordering nodes.
The $2f+1$ blocks are necessary because frontends do not verify signatures. However, this number guarantees a minimum of $f+1$ valid signatures to peers and clients.\footnote{If the frontends are programmed to perform signature verification, only $f+1$ matching blocks suffice.}
Frontends are part of the peer trust domain and are responsible for (1) relaying the envelope to the ordering cluster on behalf of the client, and (2) receiving the blocks generated by the ordering cluster and relaying them to the peers responsible for maintaining the distributed ledger.
%A \smart ordering nodes can be deployed within a peer trust domain, as long as a peer's trusted domain contains at most one of these nodes.

\subsection{Architecture}

\begin{figure}[!t]
\centering
\includegraphics[width=\columnwidth]{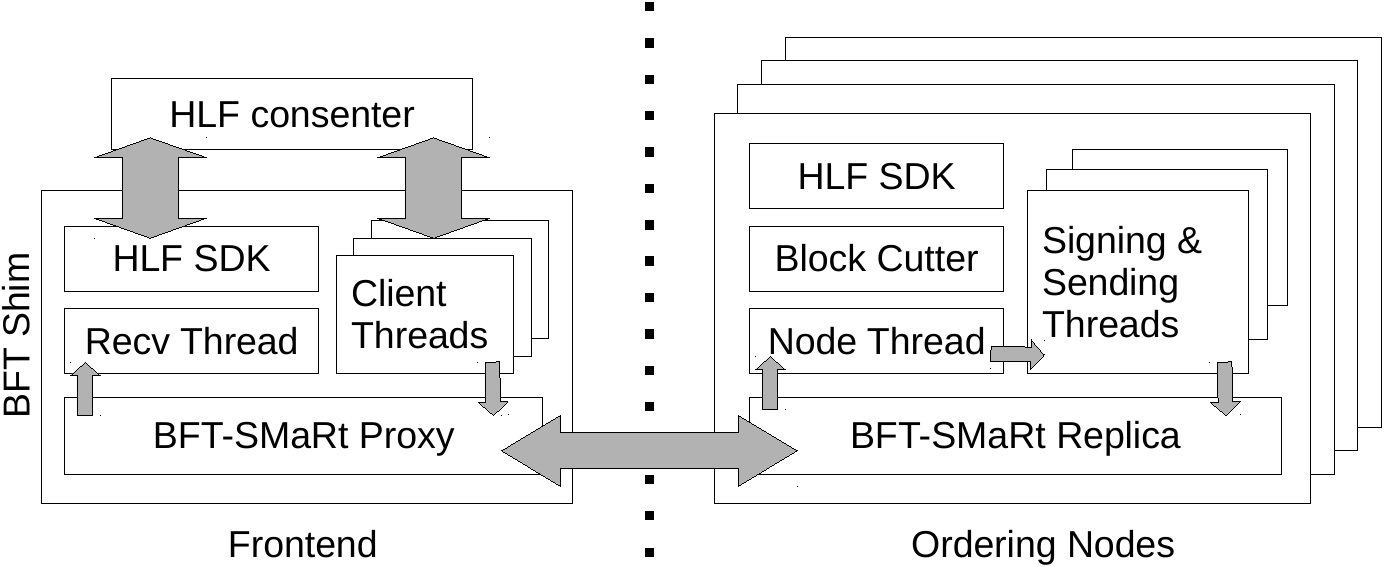}
\caption{BFT-SMaRt ordering service architecture.}
\label{fig:orderer-architecture}
\end{figure}

BFT-SMaRt's ordering service architecture is illustrated in Figure \ref{fig:orderer-architecture}.
The frontend is composed by the HLF consenter and a BFT shim.
The HLF consenter is implemented in Go and provides an interface for the HLF codebase to submit envelopes.
These envelopes are relayed to the BFT shim using sockets.
This shim is implemented in Java and maintains (1) a client thread pool that receive envelopes from the consenter and relays them to the ordering cluster, and (2) a receiver thread that collects blocks from the cluster.
Envelopes (resp. blocks) are sent to (resp. received from) the cluster through the BFT-SMaRt proxy.
The proxy does that by issuing an asynchronous invocation request to the \smart client-side library, ensuring it does not block waiting for replies.
To ensure that both the consenter and shim perform computations on equivalent data structures, the shim uses the the Hyperledger Fabric SDK to parse and assemble data structures used in HLF.

The ordering nodes are implemented on top of the BFT-SMaRt service replica, thus receiving a stream of totally ordered envelopes.
Each node maintains an object named \emph{blockcutter}, where they store the envelopes received from the service replica.
Once the blockcutter holds a pre-determined number of envelopes (the block size), it notifies the node thread that it is time to drain its envelopes and create the next block.
After the blockcutter is drained, the envelopes are assigned a sequence number associated to their future block and submitted to the signing/sending thread pool alongside with the respective block header (containing the aforementioned sequence number and the cryptographic hashes from the previous header and the hash for the current envelopes).
Notice that this thread pool does not cause non-determinism across the nodes because (1) the block header and envelopes to be assigned to new blocks are generated sequentially within the node thread, and (2) the only structures each node needs to maintain as the application state is the block header from the previous iteration of the node thread.
Similarly to the frontend, the HLF SDK is used to correctly handle and create the data structures used by the platform.
In addition, the HLF SDK is also used to generate cryptographic hashes and ECDSA (Elliptic Curve DSA) signatures that can be validated by other components of HLF.

Once the block is created and properly signed, they are transmitted to all active frontends.
This is done by providing a custom \emph{replier} (supported by the extensible API of \smart) that instead of sending the execution result (i.e., the generated block) to the invoking client, sends it to the registered \smart clients (i.e., the frontends).

\subsection{Durability and Reconfiguration}

Besides the transaction ordering and execution, \smart also provides additional capabilities that are fundamental for practical state machine replication, such as durability (of state, in case the all ordering nodes fail) and reconfiguration (of the group of ordering nodes).
Durability in particular can lead to many inefficiencies on state machine replication systems~\cite{Bes13-durasmart}.
Fortunately, our ordering service will not be subject to most of these inefficiencies as the service state is very small: just the sequence number of the next block (a 8-byte integer) and the hash of previous block (a 24-byte byte array).\footnote{It is worth to recall that in HLF the ordering nodes are not responsible for storing the blockchain, just to generate the blocks and disseminate to other peers.}
Such small state enables the execution of frequent checkpoints with little performance degradation.
This is important for limiting the size of the SMR operation logs, as they are deleted just after a new checkpoint is stored.

A small log and checkpoint allow the addition of new nodes to the ordering cluster, as the most costly operation during a group reconfiguration is the state transfer from one of the up to date nodes to the joining node~\cite{Bessani2014}.

\section{Evaluation}
\label{sec:smart-os-eval}

In this section we describe the experiments conducted to evaluate \smart's ordering service and discuss the observed results.
Our aim here is not to evaluate the whole HLF platform, but only the ordering service, which typically is the bottleneck of the system.

\subsection{Signature Generation}
\label{sec:eval-block_sigs}

The throughput of the ordering service (i.e., the rate at which envelopes are added to the blockchain) is bounded by one of three factors: a) the rate at which envelopes are ordered by \smart for a given envelope size, number of envelopes per block and number of receivers; b) the number of blocks signed per second; or c) the size of the generated blocks.
More precisely, given an envelope size $\mathit{es}$, block sizes $\mathit{bs}$, and a number of receivers $\mathit{r}$ (i.e., the frontends of endorsing and committing peers to which the ordering nodes transmit the generated blocks), peak throughput is bounded as follows:

\begin{equation}
	\mathit{TP}^{\mathit{bs,es,r}}_{\mathit{os}} \leq min(\mathit{TP}_{\mathit{sign}} \times \mathit{bs}, \mathit{TP}^{\mathit{bs,es,r}}_{\mathit{bftsmart}})
\end{equation} 

Therefore, we start by presenting a micro-benchmark designed to evaluate the performance associated with the signature of HLF blocks.\footnote{The equation consider a block is signed only once by each ordering node, however, in HLF 1.0 sometimes a block need to be signed twice. The second signature is needed to attach the block transaction to an execution context (but details are out of the scope of this paper). If this is the case for the considered application, the $TP_{signed}$ term used in the equation must be exchanged by $\frac{TP_{signed}}{2}$.}
In particular, this micro-benchmark is aimed at exploring the impact of signature parallelization using up to 16 worker threads with blocks containing 10 envelopes of 0 bytes each.
The experiment was conducted in a Dell PowerEdge R410 machine, which possesses two quad-core 2.27 GHz Intel Xeon E5520 processor with hyper-threading (thus having 16 hardware threads) and 32 GB of memory.

\paragraph{Results.}

\begin{figure}[ht]
\centering
\includegraphics[width=0.95\columnwidth]{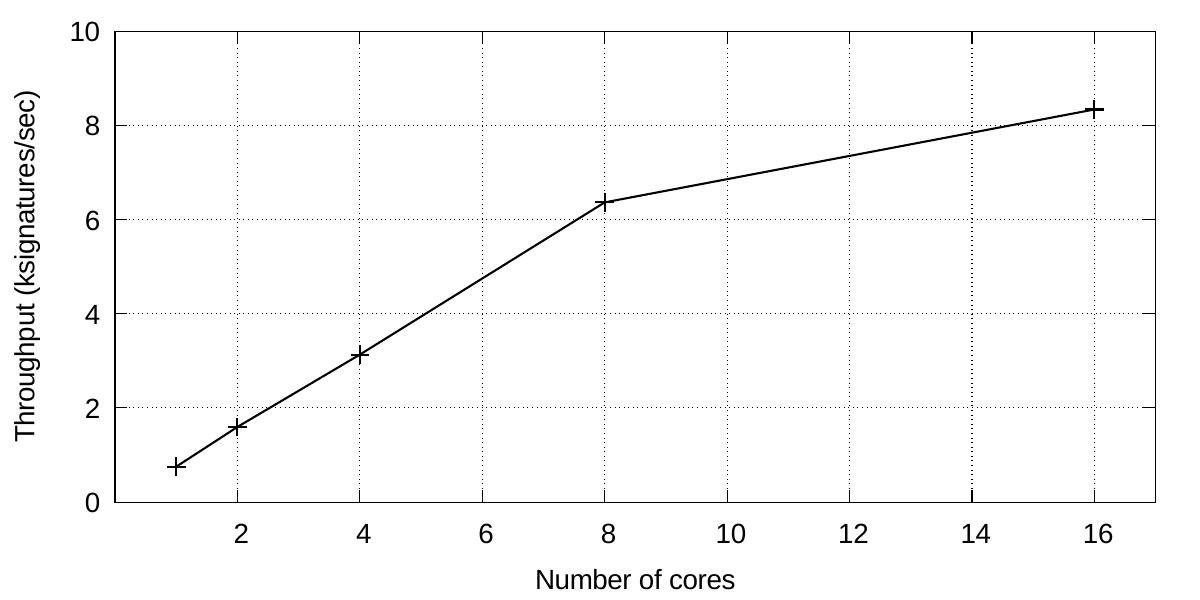}
\caption{Signature Generation for Fabric blocks.}
\label{fig:block_sigs}
\end{figure}

The results for the micro-benchmark are depicted in Figure \ref{fig:block_sigs}.
As expected, the rate of signature generation increases with the number of available worker threads, reaching a maximum rate of 8.400 ECDSA signatures/second.
This means that if each block contains 10 envelopes, we have a theoretical upper bound of 84.000 transactions/seconds in our servers for this block size.
%This value closely matches the throughput obtained in Chapter \ref{sec:smart:evaluation}, where \smart is able to reach a peak throughput of 80.000 operations/second for void requests/replies under $f=1$.

We also executed this micro-benchmarks with different envelope and block sizes, but we omit these results because they are similar to Figure \ref{fig:block_sigs}.
This is because the signatures are generated against the block header rather than against the whole block.
Since the header's size is constant regardless of the data contained in the block, the performance remains constant.
%However, since \bftsmart's throughput starts dropping after 100-bytes operations, the ordering cluster will not be able to sustain this performance.

\subsection{Ordering Cluster in a LAN}

The experiments aims to evaluate the \smart ordering service by using clients that emulate the behavior of multiple ordering service frontends.
They were executed with clusters of 4, 7, and 10 nodes, withstanding 1, 2, and 3 Byzantine faults, respectively.
Furthermore, we also fiddled with the block size, by configuring each cluster configuration to assemble blocks containing either 10 or 100 envelopes (i.e., transactions).
This is meant to observe the behaviour of each cluster when throughput is bound by either the rate of signature generation or by the rate of envelope reception.
The environment is comprised by Dell PowerEdge R410 servers connected through a Gigabit ethernet.
%More precisely, a block size of 10 envelopes will increase the rate of block creation, which in turn also increases the rate of signature generation.
%Because signature generation is parallelized, it surpasses the rate of envelope reception and consumes the entire time of the machines' CPU.
%On the other hand, a block size of 100 envelopes allows the rate of envelope reception to match the rate of signature generation and avoid CPU exhaustion.

For each micro-benchmark configured to have $x$ nodes and $y$ envelopes/block, we gathered results for (1) envelopes with different sizes, and (2) a variable number of receivers.
More precisely, each envelope size is representative of submitting to the ordering cluster: (1) a SHA-256 hash (40 bytes); (2) three ECDSA endorsement signatures (200 bytes); and (3) transaction messages of 1 and 4 kbytes.
In practice, and considering the way HLF 1.0 operates, the values related with (3) are more representative of the size of a transaction.
In particular, our limited experience shows that transactions compressed with gzip tend to be usually close to 1 kbyte.
Nonetheless, measurements for (1) and (2) are important to show the potential of the ordering service if different design choices were taken in future versions of the platform.

Measurements for the throughput associated to block generation were gathered at ordering node 0 (the leader replica of \smart's replication protocol).
To reach the system's peak throughput, each execution was performed using 16 to 32 clients distributed across 2 additional machines.
We also repeated the the micro-benchmark with 4 nodes with blocks of 100 envelopes.
All experiments used 16 signing threads (to match the number of available cores) and were repeated 3 times taking 5 minutes each.

\paragraph{Results.}

\begin{figure*}[!h] 
\centering
    \begin{subfigure}{\columnwidth}
        \includegraphics[width=\columnwidth]{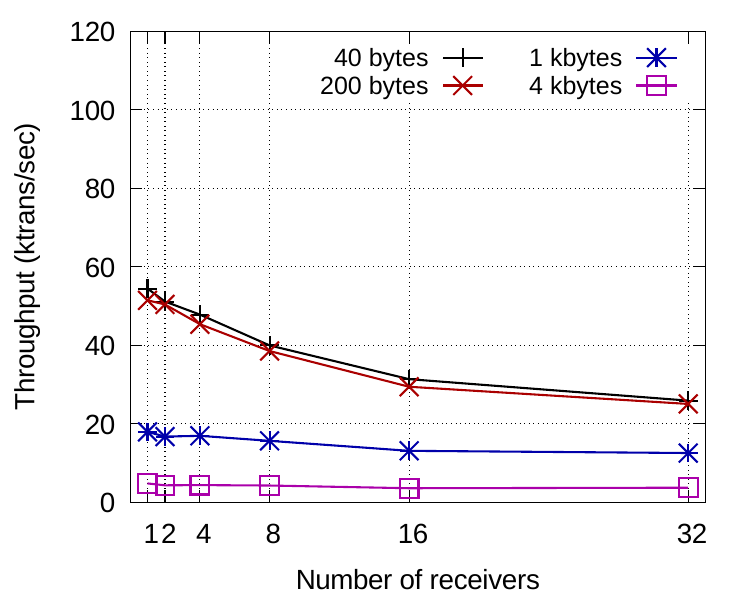}
				\caption{4 orderers, 10 envelopes/block.}
        \label{fig:receivers_f_1}
    \end{subfigure}%
		\begin{subfigure}{\columnwidth}
        \includegraphics[width=\columnwidth]{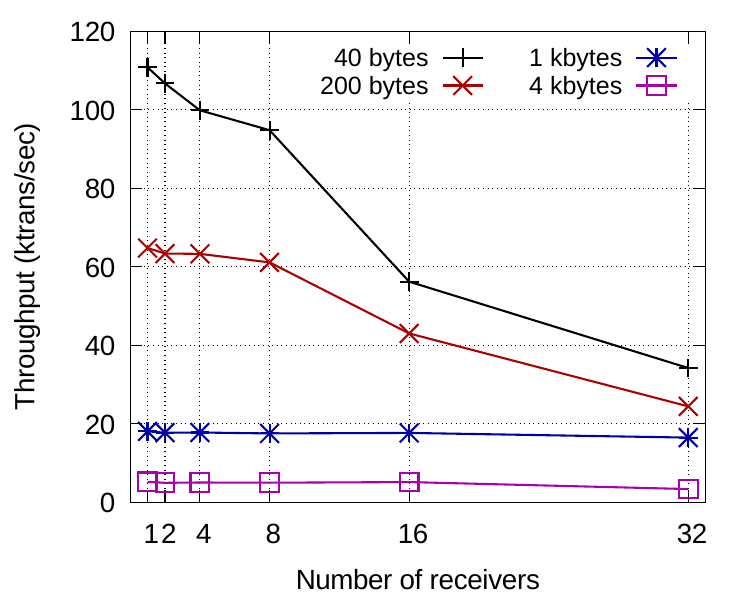}
				\caption{4 orderers, 100 envelopes/block.}
        \label{fig:receivers_f_1_k_100}
    \end{subfigure}
		
    \begin{subfigure}{\columnwidth}
        \includegraphics[width=\columnwidth]{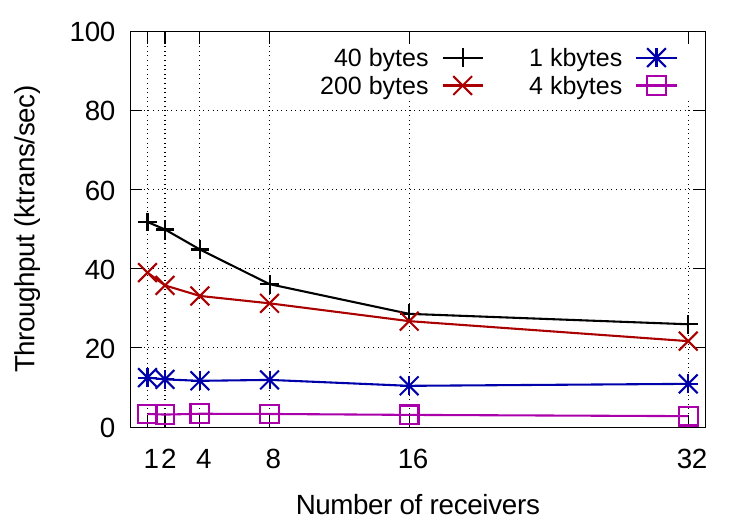}
				\caption{7 orderers, 10 envelopes/block.}
        \label{fig:receivers_f_2}    
    \end{subfigure}%
		\begin{subfigure}{\columnwidth}
        \includegraphics[width=\columnwidth]{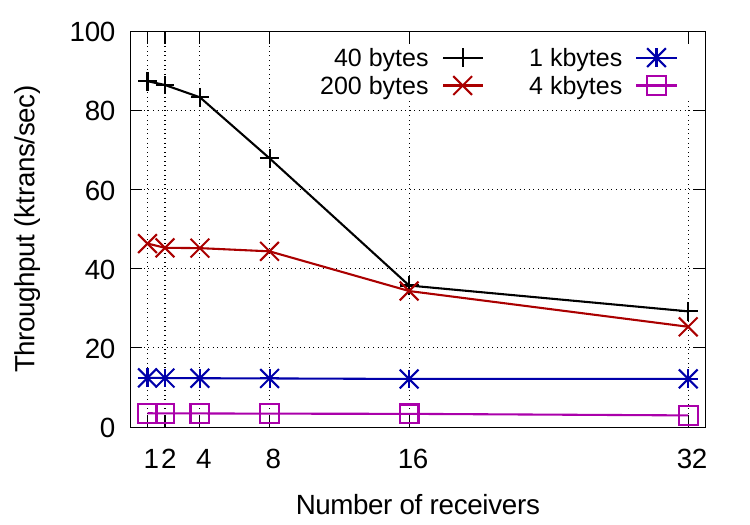}
				\caption{7 orderers, 100 envelopes/block.}
        \label{fig:receivers_f_2_k_100}    
    \end{subfigure} 
		
		\begin{subfigure}{\columnwidth}
        \includegraphics[width=\columnwidth]{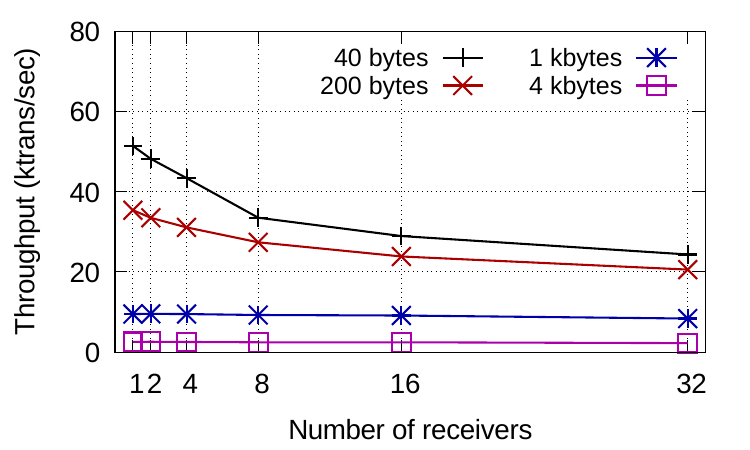}
				\caption{10 orderers, 10 envelopes/block.}
        \label{fig:receivers_f_3}    
    \end{subfigure}%
		\begin{subfigure}{\columnwidth}
        \includegraphics[width=\columnwidth]{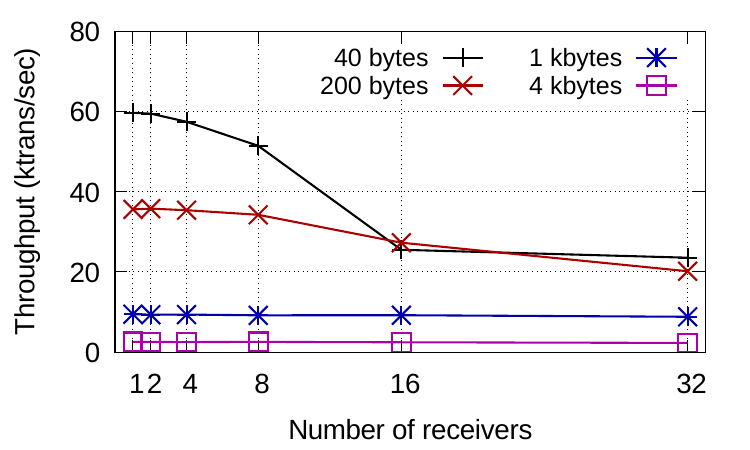}
				\caption{10 orderers, 100 envelopes/block.}
        \label{fig:receivers_f_3_k_100}    
    \end{subfigure}
		
	\caption{\smart Ordering Service throughput for different envelope, block and cluster sizes.}
		\label{fig:receivers}

\end{figure*}

The obtained results for local-area are presented in Figure \ref{fig:receivers}.
Even though throughput drops when increasing the number of receivers, the impact of the number of receivers is considerably smaller for larger transactions (1k and 4 kbytes).
This is because for these envelope sizes, the overhead of the replication protocol is greater than the overhead of transmitting blocks of 10 and 40 kbytes.
In particular, since the batch limit of the \smart is set to 400 requests, the PROPOSE message of the underlying replication protocol can have up to 0.39/1.6 Mbytes for these envelope sizes.
%Therefore, even though the observed throughput is lower for these transaction sizes (less than 2K-5K transactions/second), this scenario scales better in function of the number of replies than for transactions of 40 and 200 bytes. % because there is less contention among signing threads for I/O resources to transmit blocks to receivers. % -- as well as less memory pressure submitted to the Netty framework used by \smart.

It can be observed that when using 10 envelopes/block (Figures \ref{fig:receivers_f_1}, \ref{fig:receivers_f_2}, and \ref{fig:receivers_f_3}), the maximum throughput observed is approximately 50.000 transactions/second (when there exists only 1 to 2 receivers in the system), which is below the values observed in Section \ref{sec:eval-block_sigs}.
%Nonetheless this is to be expected, because the ordering nodes are required to sign each block, thus bounding the performance by the rate of signatures that the system is able to reach.
Nonetheless, this can be explained by the fact that signature generation needs to share CPU power with the replication protocol, hence creating a thug-of-war between the application's worker threads and \smart's I/O threads and queues -- in particular, \smart alone can take up to 60\% of CPU usage when executing a void service with asynchronous clients.
Hence, the performance drop in relation to the micro-benchmark from Section \ref{sec:eval-block_sigs} -- which executed in a single machine, stripped of the overhead associated with \smart -- is to be expected.
Moreover, for up to 2 receivers and envelope sizes of 1 and 4 kbytes, the peak throughput becomes similar to the results observed in \cite{Bessani2014}.
This is because for these request sizes \smart is unable to order envelopes at a rate equal to the rate at which the system is able to produce signatures.

Figures \ref{fig:receivers_f_1_k_100}, \ref{fig:receivers_f_2_k_100}, and \ref{fig:receivers_f_3_k_100} show the results obtained for 100 envelopes/block, when each node is not subject to CPU exhaustion.
It can be observed that, across all cluster sizes, throughput is significantly higher for smaller envelope sizes and up to 8 receivers.
%In particular, for an envelope size of 40 bytes with a single receiver the throughput doubles.
This happens because even though each node creates blocks at a lower rate -- approximately 1100 blocks per seconds -- each block contains 100 envelopes instead of only 10. %\footnote{The reason why this behavior contradicts what is reported in Section \ref{sec:eval-block_sigs} is because the rate at which nodes generate signed blocks is directly dependent on the rate at which envelopes are drained from the blockcutter.}
Moreover, this configuration makes the rate at which envelopes are ordered to become similar to the rate at which blocks are created.
This means that for smaller envelope sizes, it is better to adjust the nodes' configuration to avoid consuming all the CPU time and rely on the rate of envelope arrival.
However, for envelopes of 1 and 4 kbytes the behavior is similar to using 10 envelopes/block, specially from 7 nodes onward.
This is because for larger envelope sizes -- as discussed previously -- the predominant overhead becomes the replication protocol.
Interestingly, for a larger number of receivers (16 and 32), throughput converges to similar values across all combinations of envelope/cluster/block sizes.
Whereas for larger envelope sizes this is due to the overhead of the replication protocol, for smaller envelope sizes this happens because the transmission of blocks to the receivers becomes the predominant overhead.%, whereas for larger envelope sizes -- as discussed previously -- the predominant overhead becomes the ordering protocol.

\subsection{Geo-distributed Ordering Cluster}

In addition to the aforementioned micro-benchmarks deployed in a local datacenter, we also conducted a geo-distributed benchmark focused on collecting latency measurements at 4 frontends scattered across the Americas, with the nodes of the ordering service distributed all around the world: Oregon, Ireland, Sydney, and S\~{a}o Paulo (four \smart replicas), with Virginia standing as \wheat's additional replica (five replicas).
Since signatures generation requires considerable CPU power, we used instances of the type \emph{m4.4xlarge}, with 16 virtual CPUs each.
The frontends were deployed in Canada (frontend only), Oregon (collocated with leader node weighting $\mathit{V_{max}}$ in \wheat), Virginia (collocated with non-leader node, but still weighting $\mathit{V_{max}}$) and S\~{a}o Paulo.\footnote{According to the \wheat binary weight distribution for BFT state machine replication~\cite{Sousa2015}, when using five replicas, two of them will have weight $\mathit{V_{max}} = 2$ and the remaining three will have $\mathit{V_{min}} = 1$.}
Each frontend was configured to launch enough client threads to keep node throughput always above 1000 transactions/second.

\paragraph{Results.} 

\begin{figure*}[t] 
    \begin{subfigure}{\columnwidth}
    	\centering
        \includegraphics[scale=0.28]{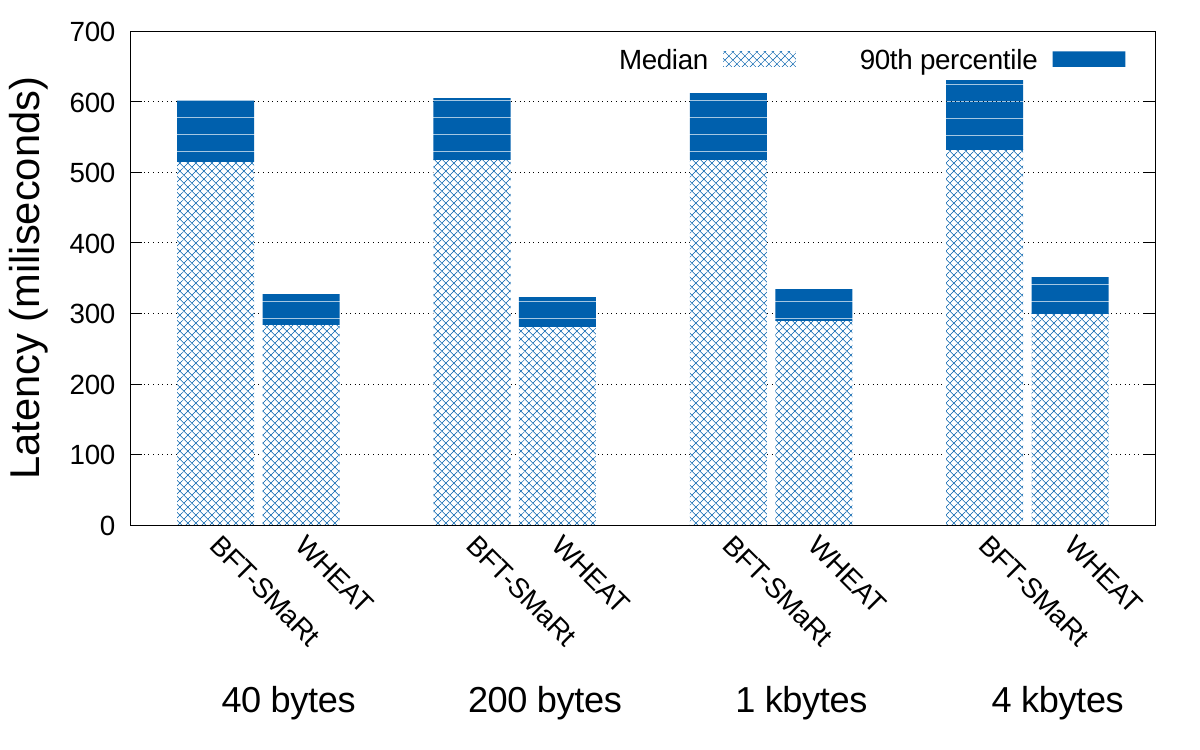}
				\caption{Canada (clients only).}
        \label{fig:hlf-canada10}
    \end{subfigure}%
    \begin{subfigure}{\columnwidth}
    	\centering
        \includegraphics[scale=0.28]{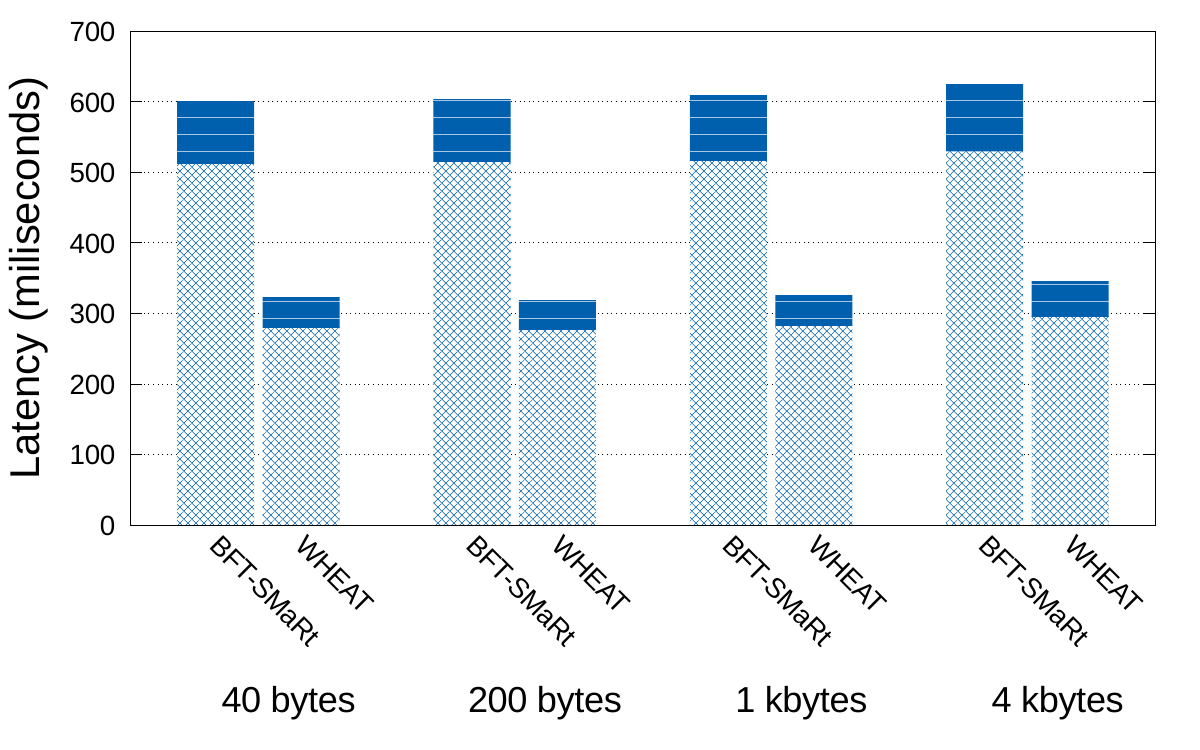}
				\caption{Oregon (weighted $\mathit{V_{max}}$, leader node).}
        \label{fig:hlf-oregon10}    
    \end{subfigure} 		
    \begin{subfigure}{\columnwidth}
    	\centering
        \includegraphics[scale=0.28]{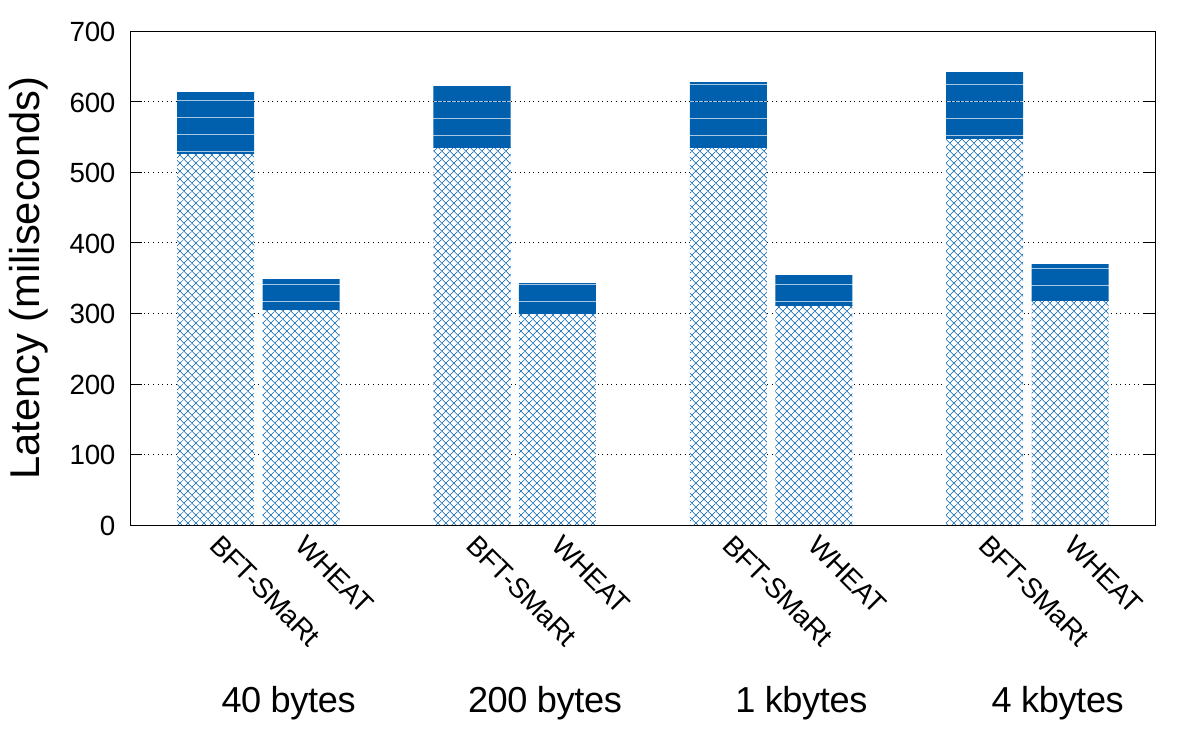}
				\caption{Virginia (weighted $\mathit{V_{max}}$).}
        \label{fig:hlf-virginia10}
    \end{subfigure}%
    \begin{subfigure}{\columnwidth}
    	\centering
        \includegraphics[scale=0.28]{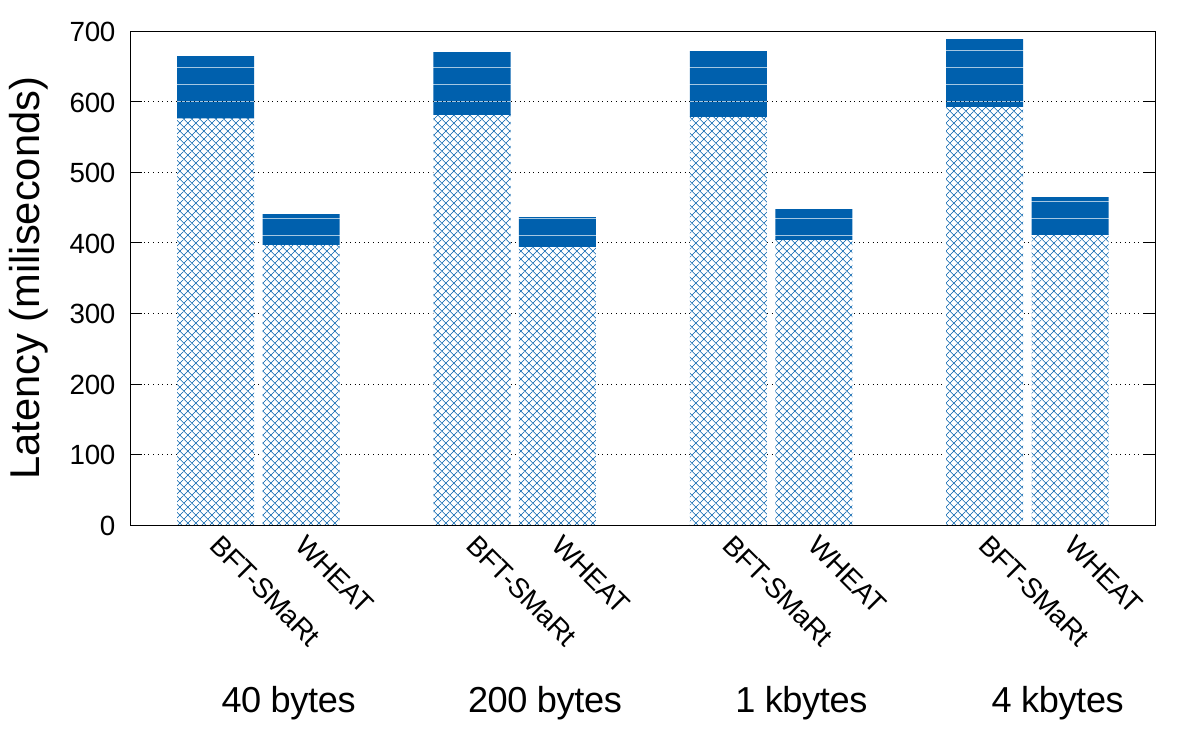}
				\caption{S\~{a}o Paulo (weighted $\mathit{V_{min}}$).}
        \label{fig:hlf-saopaulo10}    
    \end{subfigure} 
		
    \caption{Amazon EC2 latency results (4 receivers, blocks with 10 envelopes).}
		\label{fig:hlf-wan10}
\end{figure*}

Figure \ref{fig:hlf-wan10} presents the results for the geo-distributed micro-benchmark with a a block size of 10 envelopes.
As expected, \wheat's latency is consistently lower than \smart's across all frontends by almost 50\%.
It is worth pointing out that envelope size has a relatively minor impact on latency: across all regions, the difference between a 40 and a 4k bytes envelope was never above 29 milliseconds for any percentile or protocol. %(-4\%).
In fact, it is the placement of the frontends that can exhibit a larger impact on latency: the difference between Virginia (weighted $\mathit{V_{max}}$) and S\~{a}o Paulo (weighted $\mathit{V_{min}}$) is above 43 milliseconds for \smart (+6.5\%) and above 90 milliseconds (+23\%) for \wheat.
In addition, the difference between S\~{a}o Paulo and Oregon/Canada is even larger (58 milliseconds for \smart and 100 miliseconds for \wheat, corresponding to an increase of  +8,5\% and +27\% respectively).

\begin{figure*}[!t] 
    \begin{subfigure}{\columnwidth}
    	\centering
        \includegraphics[scale=0.28]{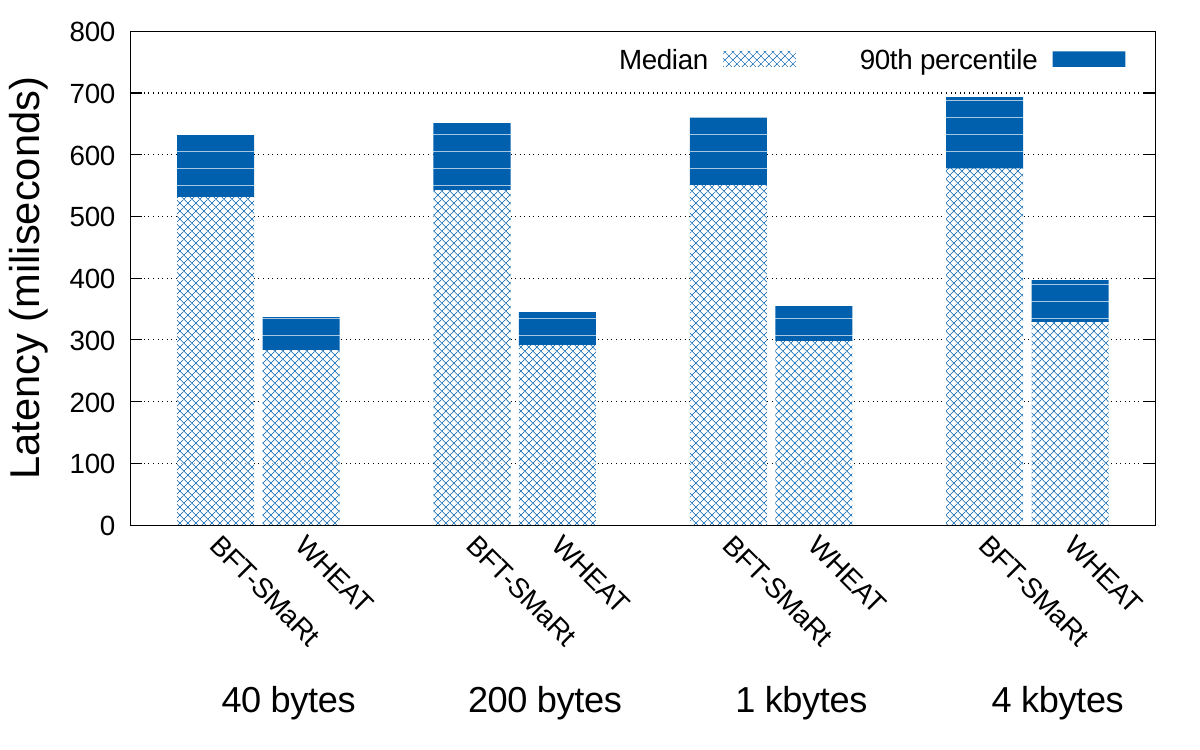}
				\caption{Canada (clients only).}
        \label{fig:hlf-canada100}
    \end{subfigure}%
    \begin{subfigure}{\columnwidth}
    	\centering
        \includegraphics[scale=0.28]{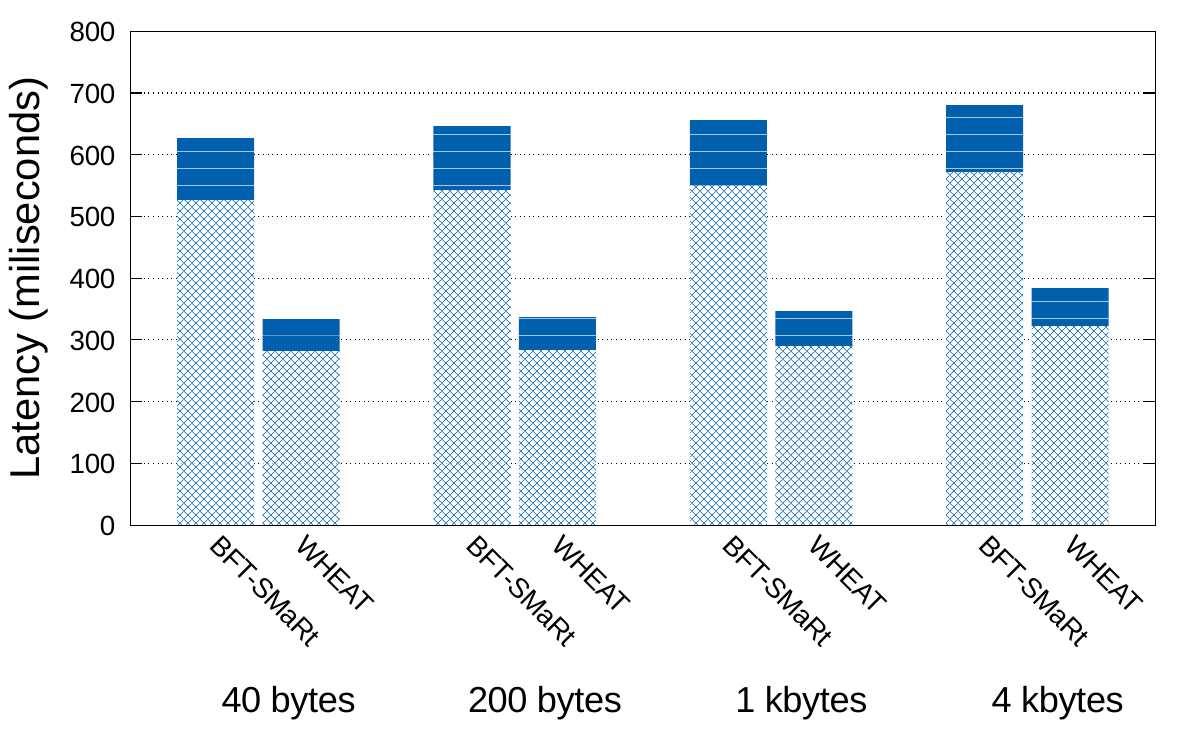}
				\caption{Oregon (weighted $\mathit{V_{max}}$, leader node).}
        \label{fig:hlf-oregon100}    
    \end{subfigure} 
    \begin{subfigure}{\columnwidth}
    	\centering
        \includegraphics[scale=0.28]{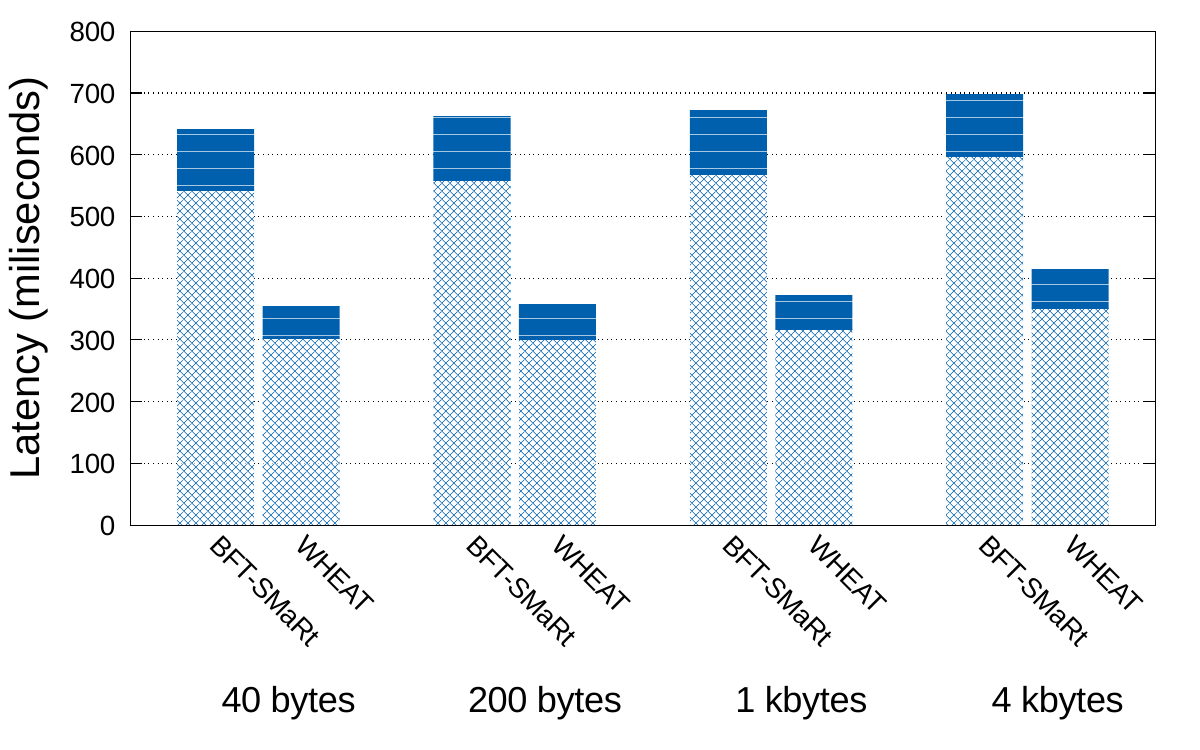}
				\caption{Virginia (weighted $\mathit{V_{max}}$).}
        \label{fig:hlf-virginia100}
    \end{subfigure}%
    \begin{subfigure}{\columnwidth}
    	\centering
        \includegraphics[scale=0.28]{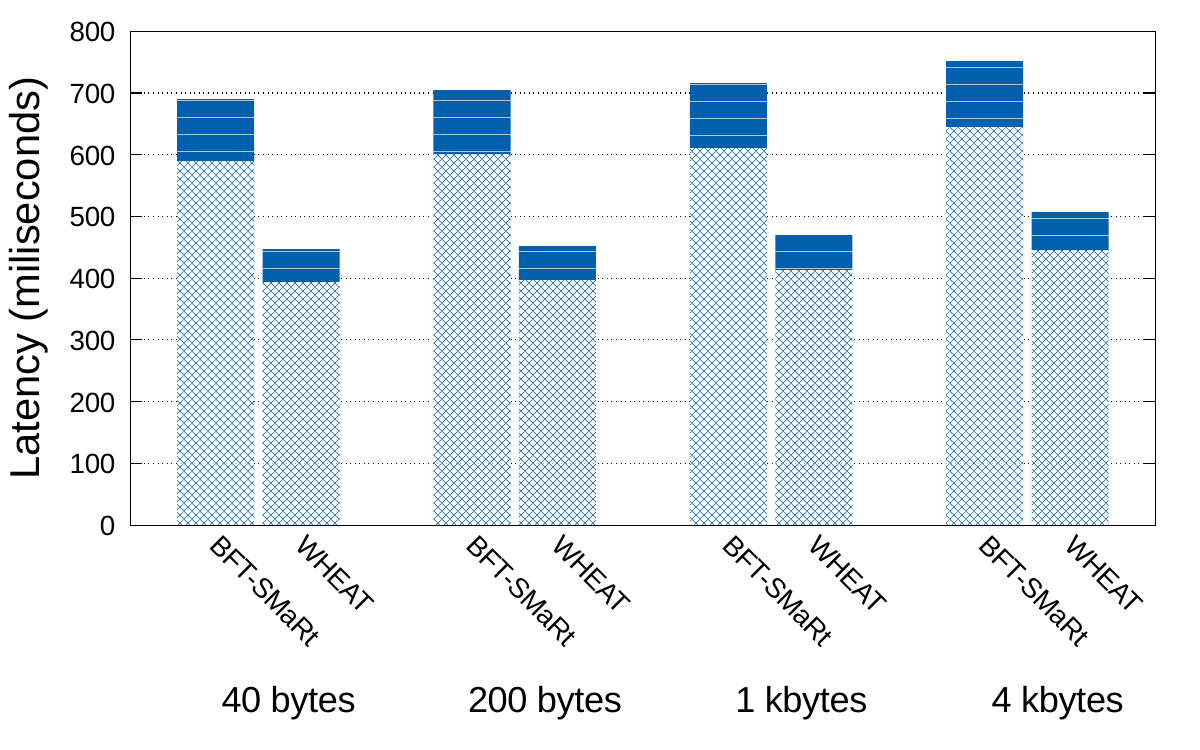}
				\caption{S\~{a}o Paulo (weighted $\mathit{V_{min}}$).}
        \label{fig:hlf-saopaulo100}    
    \end{subfigure} 		
    \caption{Amazon EC2 latency results (4 receivers, blocks with 100 envelopes).}
    \label{fig:hlf-wan100}
\end{figure*}

We also repeated the experiment for blocks of 100 envelopes (Figure \ref{fig:hlf-wan100}).
The results are similar to the previous experiment, but with increased latency (up to 63 milliseconds higher).
This is because with similar workload but a larger block size, the rate of block generation decreases, which has a direct impact on latency.

\section{Related Work}
\label{sec:rel-work}

The concept of blockchain was originally introduced by Bitcoin to solve the double spending problem associated with crypto-currency in permissionless peer-to-peer networks~\cite{Nakamoto_bitcoin:a}.
Since Bitcoin's inception and widespread adoption, other platforms based on Proof-of-Work blockchain have emerged.
Within these new platforms, Ethereum is particularly relevant for its support of smart contracts \cite{Wood15}.
%Both Bitcoin and Ethereum are permissionless platforms.

Because of the known performance penalty associated with Proof-of-Work creation and the fact that Blockchain technology is gaining the attention of many industries, the idea of permissioned blockchains are quickly gaining traction.
Examples of other permissioned blockchain platforms include Chain,\footnote{\url{https://chain.com/}} which uses the Federated Consensus algorithm.\footnote{\url{https://chain.com/docs/1.2/protocol/papers/federated-consensus}}
Tendermint \cite{Kwon2016} implements the BFT protocol designed by Buchman et. al.\cite{Buchman2016}.
Kadena \cite{Martino2016} uses a variant of the Raft consensu protocol \cite{Ongaro2014} adapted to Byzantine faults \cite{Copeland2014TangaroaAB}.
Finally, Symbiont Assembly\footnote{\url{https://symbiont.io/technology/assembly/}} uses a Go implementation of the Mod-SMaRt algorithm \cite{Sousa12:::} and heavily follows the design of \smart.
A recent survey compares all these permissioned protocols and points \smart as a prominent candidate for implementing this type of ledgers.

\section{Conclusion} 
\label{sec:conclusions}

The evaluation confirms the initial hypothesis about peak throughput being bounded either by the rate at which signatures can generated by a replica, or the rate of envelopes ordered by the total order protocol.
Moreover, the results also suggest that, for smaller envelope sizes, increasing the block size while decreasing the rate of signature generation can yield higher transactional throughput than to simply rely on the maximum possible rate of signature generation.
Nonetheless, for a higher number of repliers, throughput tends to converge to similar values across all micro-benchmarks.
Even when transmitting blocks of 400 kbytes to 32 receivers in a cluster of 10 nodes, the ordering service still reaches a peak throughput of approximately 2200 transactions/second -- which is more $2\times$ of Ethereum's theoretical peak of 1000 transactions/second \cite{Buterin2016}, and vastly superior than Bitcoin's peak of 7 transaction/second \cite{Vukolic2016}.
Finally, latency measurements taken from a geo-replicated setting are also shown attractive, with values within half a second under moderate workload using \wheat, even when accounting for large block sizes.

\paragraph{Acknowledgements.} This work was supported by an IBM Faculty Award, by FCT through projects LaSIGE (UID/CEC/00408/2013) and IRCoC (PTDC/EEI-SCR/6970/2014), and by the European Commission through the H2020 SUPERCLOUD project (643964).

{\small
\bibliographystyle{plain}
\bibliography{References/myrefs}
}

\end{document}